\documentclass[10pt, conference]{IEEEtran}
\IEEEoverridecommandlockouts
\usepackage{cite}
\usepackage{verbatim}
\usepackage{url}
\usepackage{grffile}
\usepackage{graphicx}
\usepackage{subcaption}
\usepackage{amsmath,amssymb,amsfonts}
\usepackage[detect-all]{siunitx}
\usepackage[hidelinks]{hyperref}
\usepackage{cleveref}
\usepackage{tikz}
\usepackage{algorithmic}
\usepackage{epstopdf}
\usepackage[utf8]{inputenc}

\def\BibTeX{{\rm B\kern-.05em{\sc i\kern-.025em b}\kern-.08em
    T\kern-.1667em\lower.7ex\hbox{E}\kern-.125emX}}

\newcommand\copyrighttext{%
  \footnotesize \textcopyright 2023 IEEE. Personal use of this material is permitted. Permission from IEEE must be obtained for all other uses, in any current or future media, including reprinting/republishing this material for advertising or promotional purposes, creating new collective works, for resale or redistribution to servers or lists, or reuse of any copyrighted component of this work in other works.}

\newcommand\copyrightnotice{%
\begin{tikzpicture}[remember picture,overlay]
\node[anchor=south,yshift=10pt] at (current page.south) {\fbox{\parbox{\dimexpr\textwidth-\fboxsep-\fboxrule\relax}{\copyrighttext}}};
\end{tikzpicture}%
}    

\usepackage[]{footmisc}

\begin{document}

\title{On the Mobility Analysis of UE-Side Beamforming for Multi-Panel User Equipment in 5G-Advanced\\
}

\author{
    \IEEEauthorblockN{
        Subhyal Bin Iqbal\IEEEauthorrefmark{1}\IEEEauthorrefmark{2}, Salman Nadaf\IEEEauthorrefmark{1}\IEEEauthorrefmark{3},  Umur Karabulut\IEEEauthorrefmark{1}, Philipp Schulz\IEEEauthorrefmark{2}, Anna Prado \IEEEauthorrefmark{3}, \\ Gerhard P. Fettweis\IEEEauthorrefmark{2} and Wolfgang Kellerer \IEEEauthorrefmark{3} 
    }
    \IEEEauthorblockA{\IEEEauthorrefmark{1} Nokia Standardization and Research Lab, Munich, Germany}
    \IEEEauthorblockA{\IEEEauthorrefmark{2} Vodafone Chair for Mobile Communications Systems, Technische Universität Dresden, Germany}
    \IEEEauthorblockA{\IEEEauthorrefmark{3} Chair of Communication Networks, Technical University of Munich, Germany}
}

\maketitle

%
\copyrightnotice

\begin{abstract}
Frequency range 2 (FR2) has become an integral part of 5G networks to fulfill the ever-increasing demand for data hungry-applications. However, radio signals in FR2 experience high path and diffraction loss, which also pronounces the problem of inter and intra-cell interference. As a result, both the serving and target links are affected, leading to radio link failures (RLFs) and handover failures (HOFs), respectively. To address this issue, multi-panel user equipment (MPUE) is proposed for 5G-Advanced whereby multiple spatially distinct antenna panels are integrated into the UE to leverage gains from antenna directivity. It also opens the possibility of using UE-side Rx-beamforming for each panel. In this paper, three different Rx-beamforming approaches are proposed to improve the serving link, the target link, and the handover process for an MPUE equipped with three directional panels. Thereafter, the mobility performance is analyzed in a system-level simulation for a multi-beam FR2 network. Results have shown that the proposed schemes can help reduce RLFs by 53\% and HOFs by 90\%.
\end{abstract}

\begin{IEEEkeywords}
frequency range 2 (FR2), radio link failures (RLFs), handover failures (HOFs), multi-panel user equipment (MPUE), 5G-Advanced, UE-side Rx-beamforming, mobility performance, system-level simulation.
\end{IEEEkeywords}


%
\vspace{-1mm}
\section{Introduction} \label{Section1}

Although the reliance of 5G multi-beam networks on the large spectrum availability offered by frequency range 2 (FR2) offers a practical solution to fulfill the demand of data-hungry applications, it introduces additional challenges to the link budget such as higher free-space path loss and penetration loss \cite{b1}. This leads to rapid degradation of the signal power in mobile environments where there are many static and moving obstacles and makes handovers (HOs) from the serving to target cells more challenging. Furthermore, it also pronounces the problem of both inter and intra-cell interference \cite{b2}. Consequently, this affects the serving link that exists between the UE and the serving cell, leading to radio link failures (RLFs). It also affects the target link that exists between the UE and the target cell, leading to handover failures (HOFs). In both these types of mobility failures, the UE would be prevented from performing a HO to another target cell with a better link, leading to an outage in the network.

On the UE architecture side, multi-panel UEs (MPUEs) \cite{b3, b4} offer a solution to this problem by integrating multiple spatially distant antenna panels into the UE, thus offering both directional gain and inter-cell interference suppression from neighboring cells. In this paper, we propose a solution based on the MPUE architecture with UE-side receiver (Rx)-beamforming that individually caters to improving both the serving and target links and on making the handover process more reliable by integrating Rx-beamformed measurements into the handover process. The Rx-beamforming is based on 3GPP's beamforming framework \cite[Section 6.1]{b5} where Rx beams are swept for a given fixed transmit (Tx) beam and a narrow refined beam is selected to give the beam-pair for communication between the UE and the serving cell.

Both MPUEs and UE-side Rx-beamforming are an essential part of 5G-Advanced \cite{b6} and this paper offers a detailed system-level mobility performance outlook into the benefits of Rx-beamforming and the different ways in which it can reduce outage by improving the serving and target links and making the handover process more reliable. The authors in \cite{b61} consider Rx-beamforming for MPUEs but the focus is at studying the impact of hand blockages on beam management, whereas inter-cell mobility performance is not considered. In \cite{b62} the authors consider Rx-beamforming in a mobile environment with a single base station (BS) but do not consider MPUEs. The novelty of this paper is three-fold. Firstly, to the best of our knowledge, the system-level mobility performance for MPUEs with Rx-beamforming-centric serving link improvement has not been investigated in the literature before. Secondly, there has not been any mobility study for MPUEs where Rx-beamforming has been integrated into the layer 3 (L3) measurements-based handover decision-making process. Thirdly, a novel enhancement proposed in \cite{b65} to acquire the narrow refined beam of the target cell before a handover in order to improve the target link is validated and analyzed. 

The rest of the paper is organized as follows. In \Cref{Section2}, we provide insights into the inter-cell and intra-cell mobility procedures and the signal-to-noise-plus-interference ratio (SINR) model. In  \Cref{Section3}, we explain the UE-side beamforming in terms of the MPUE architecture and the different approaches whereby it is integrated into the SINR and handover models. The simulation scenario used in the performance evaluation is discussed in \Cref{Section4}. Then in \Cref{Section5}, the mobility key performance indicators (KPIs) are presented and the mobility performance of UE-side beamforming with different approaches is compared with non-beamformed MPUEs. Finally, in \Cref{Section6}, we conclude the paper and provide an outlook for future enhancements.

\section{Network Model} \label{Section2}
In this section, the inter-cell and intra-cell mobility that form part of the handover and beam management procedures, respectively, are reviewed along with the SINR model. 

\subsection{Inter-cell Mobility} \label{Subsection2.1}
Inter-cell mobility relates to HOs between cells in the network. A pre-requisite for a  successful HO from the serving cell to the target cell is that the physical layer reference signal received power (RSRP) measurements undergo filtering to mitigate the effect of channel impairments. The HO model that is considered in this paper is the baseline HO mechanism of \textit{3GPP Release 15} \cite{b7, b8}. In the 5G multi-beam network, each UE has the capability to measure the raw RSRP values $P_{c,b}^\textrm{RSRP}(n)$ (in dBm) at a discrete time instant $n$ from each Tx beam $b \in B$ of cell $c \in C$, using the synchronization signal block (SSB) bursts that are periodically transmitted by the BS. The separation between the time instants is denoted by $\Delta t$ ms. At the UE end, L1 and L3 filtering are then applied sequentially to the raw RSRPs in order to mitigate the effects of fast-fading and measurement errors and determine the L3 cell quality of the serving and neighboring cells. L1 filtering implementation has not been specified in 3GPP standardization and is UE-specific. Herein, we use a moving average filter for L1 filtering, where the L1 filter output is expressed as
\vspace{-0.2\baselineskip}
\begin{equation}
\label{Eq1}
    P_{c,b}^\textrm{L1}(m) = \frac{1}{N_\mathrm{L1}} \sum_{i=0}^{N_\mathrm{L1}-1}P_{c,b}^\textrm{RSRP}(m-\omega i), \ m = n\omega 
\end{equation}
where $\omega \in \mathbb{N}$ is the L1 measurement period (aligned with the SSB periodicity) that is normalized by the time step duration~$\Delta t$, and $N_\mathrm{L1}$ is the number of samples that are averaged in each L1 measurement period. The L1 beam measurements are then used for cell quality derivation, where we first consider the set $B_{\mathrm{str},c}$ of strongest beams with signal measurements that are above a certain threshold $P_{\mathrm{thr}}$. $B_{\mathrm{str},c}$ is, thus, defined as
\vspace{-0.2\baselineskip}
\begin{equation}
\label{Eq2}
    B_{\mathrm{str},c}(m) = \big\{b \  | \ P_{c,b}^\textrm{L1}(m) > P_{\mathrm{thr}} \big\}.
\end{equation}
After this, up to $N_\mathrm{str}$ beams representing the subset $B_{\mathrm{str},c}^{\prime}$ of $B_{\mathrm{str},c}$ with the strongest  $P_{c,b}^\textrm{L1}(m)$ are taken and averaged to derive the L1 cell quality of cell $c$ as 
\begin{equation}
\vspace{-0.2\baselineskip}
\label{Eq3}
    P_{c}^\textrm{L1}(m) = \frac{1}{|B_{\mathrm{str},c}^{\prime}|} \sum_{b \in B_{\mathrm{str},c}^{\prime}}  P_{c,b}^\textrm{L1}(m).
\end{equation}
The cardinality of the set is given by $|\cdot|$ and the set $B_{\mathrm{str},c}(m)$ is taken as $B_{\mathrm{str},c}^{\prime}$ if $|B_{\mathrm{str},c}(m)| < N_\mathrm{str}$. If $B_{\mathrm{str},c}(m)$ is empty, the highest $P_{c,b}^\textrm{L1}(m)$ is taken as the L1 cell quality $P_{c}^\textrm{L1}(m)$.

The L1 cell quality is further smoothed by L3 filtering to yield the L3 cell quality. An infinite impulse response (IIR) filter is used for L3 filtering, where the L3 filter output is expressed as
\vspace{-0.2\baselineskip}
\begin{equation}
\label{Eq4}
    P_{c}^\textrm{L3}(m) = \alpha_c  P_{c}^\textrm{L1}(m) + (1-\alpha_c)P_{c}^\textrm{L3}(m-\omega),
\end{equation}
where $\alpha_c = (\frac{1}{2})^{\frac{k}{4}}$ is the forgetting factor that controls the impact of older L3 cell quality measurements $P_{c}^\textrm{L3}(m-\omega)$ and $k$ is the filter coefficient of the IIR filter \cite{b7}.

Similarly, the L1 RSRP beam measurement $P_{c,b}^\textrm{L1}$ of each beam $b$ of cell $c$ also undergoes L3 filtering, where the output is now the L3 beam measurement $P_{c,b}^\textrm{L3}$ 

\vspace{-0.6\baselineskip}
\begin{equation}
\label{Eq5}
    P_{c,b}^\textrm{L3}(m) = \alpha_b  P_{c,b}^\textrm{L1}(m) + (1-\alpha_b)P_{c,b}^\textrm{L3}(m-\omega),
\end{equation}
where $\alpha_b$ can be configured independently of $\alpha_c$.

L3 cell quality $P_{c}^\textrm{L3}(m)$ is an indicator of the average downlink signal strength for a link that exists between a UE and cell~$c$. It is used by the network to trigger the HO from the serving cell $c_0$ to one of its neighboring cells, termed as the target cell $c^{\prime}$. For intra-frequency HO decisions, typically the A3 trigger condition is configured for measurement reporting~\cite{b7}. The UE is triggered to report the L3 cell quality measurement of the target cell $P_{c^{\prime}}^\textrm{L3}(m)$ and L3 beam measurements $P_{c^{\prime},b}^\textrm{L3}(m)$ to its serving cell $c_0$ when the A3 trigger condition, i.e.,
\vspace{-0.2\baselineskip}
\begin{equation}
\label{Eq6}
     P_{c_0}^\textrm{L3}(m) + o^\mathrm{A3}_{c_0, c^{\prime}} < P_{c^{\prime}}^\textrm{L3}(m) \ \text{for} \  m_0 - T_\mathrm{TTT,A3} < m < m_0,
\end{equation}
expires at the time instant $m=m_0$ for $c^{\prime} \neq c_0$, where $o^\mathrm{A3}_{c_0, c^{\prime}}$ is termed as the HO offset between cell $c_0$ and $c^{\prime}$ and the observation period in (\ref{Eq6}) is termed as the time-to-trigger $T_\mathrm{TTT,A3}$ (in ms).

Once the serving cell $c_0$ has received the L3 cell quality measurements, it sends out a HO request to the target cell~$c^{\prime}$, which is typically the strongest cell, along with the L3 beam measurements $P_{c^{\prime},b}^\textrm{L3}(m)$ of the target cell $c^{\prime}$. Thereafter, the target cell prepares contention-free random access (CFRA) resources for beams $b \in B_{\mathrm{prep},c^{\prime}}$, e.g., with the highest signal power based on the reported L3 beam measurements. The target cell replies by acknowledging the HO request and provides a HO command to the serving cell, which includes the information required by the UE to access the target cell. The serving cell forwards the HO command to the UE and once the UE receives this message, it detaches from its serving cell $c_0$ and initiates random access towards the target cell $c^{\prime}$ through a target beam $b^{\prime}$ using the CFRA resources.

\subsection{Intra-cell Mobility} \label{Subsection2.2}
Intra-cell mobility relates to a set of L1 and L2 Tx beam management procedures for the determination and update of serving Tx beam(s) for each UE within a serving cell $c_0$, as defined in \textit{3GPP Release 15} \cite{b5}. One of the key components is Tx beam selection, where the UE uses network assistance to select the serving beam $b_0$ that it uses to communicate with $c_0$ \cite[Section II-A]{b4}. This is based on the periodic reporting of selected L1 beam measurements by the UE to the network that were initially received as raw beam measurements through SSB bursts. The other key component is beam failure detection, where the aim is to detect a failure of the serving beam $b_0$ that is determined by its radio link quality SINR \cite{b10}. If a beam failure is detected, the UE is prompted to initiate a beam recovery failure (BFR) procedure where it aims to recover another beam of the serving cell $c_0$. To that effect, the UE attempts random access on the target beam $b^{\prime}$ that has the highest L1 RSRP beam measurement $P_{c_0,b}^\textrm{L1}(m)$ and then waits for the BS to send a random access response indicating that the access was successful. If the first attempt is unsuccessful, the UE attempts another random access using $b^{\prime}$. In total, $N_\mathrm{BAtt}$ such attempts are made at time intervals of $T_\mathrm{BAtt}$. If all such attempts are unsuccessful, an RLF is declared.

\subsection{SINR Model} \label{Subsection2.3}
The average downlink  SINR at the discrete time instant $m$ for Tx beam $b \in B$ of cell $c \in C$ is denoted as $\gamma_{c,b}(m)$. It is evaluated using the Monte-Carlo approximation given in \cite{b2} for the strict fair resource scheduler, where all UEs in the network get precisely the same amount of resources. As will be seen later in \Cref{Section4}, the SINRs of the serving beam-cell pair $\gamma_{c_0,b_0}(m)$ and target beam-cell pair $\gamma_{c^{\prime},b^{\prime}}(m)$ have a key role in the RLF and HOF models, respectively.

\section{UE-Side Rx-Beamforming Model}  \label{Section3}
In this section, the UE-side Rx-beamforming for the MPUE architecture is explained along with the three different UE-side beamforming approaches.

\subsection{UE-Side Beamforming with MPUE} \label{Subsection3.1}
An MPUE in $edge$ design with three integrated directional panels is considered \cite{b3, b4}. Each directional panel $d \in D$ is assumed to have four antenna elements in a 1$\times$ 4 configuration with a spacing of 0.5$\lambda$, where the wavelength $\lambda = c/f_\mathrm{FR2}$  and $f_\mathrm{FR2}$ is the FR2 carrier frequency. Thereafter, each panel is considered to have the beamforming capabilities for Rx-beamforming refinement, generating directional beams $r \in R$ for each of its panels $d \in D$, where $r \in \{1,\ldots,7\}$.

\begin{figure}[!t]
\textit{\centering
\includegraphics[width = 0.96\columnwidth]{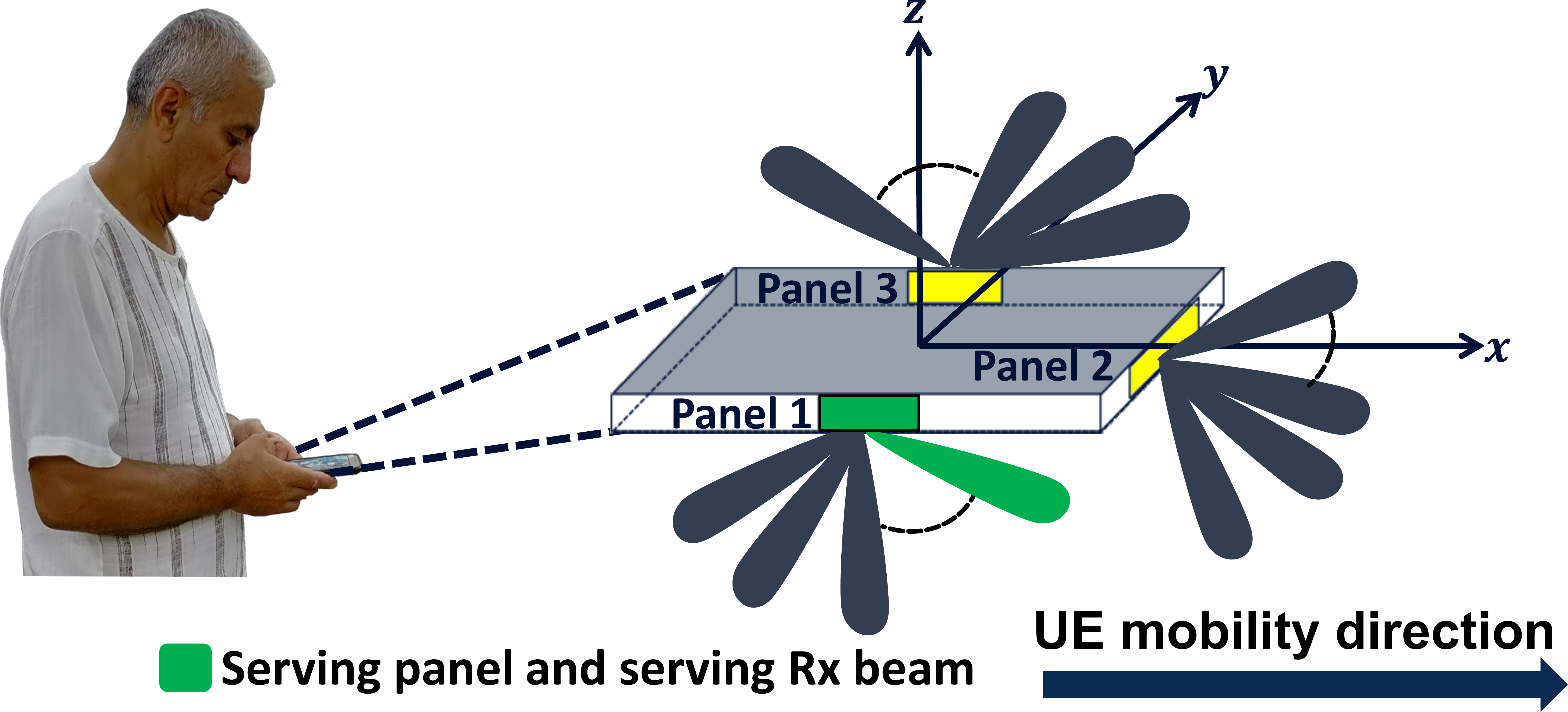}
\vspace{-4pt}
\caption{An illustration of the serving panel and serving Rx beam in the MPUE context, where the MPUE has three integrated panels in the \textit{edge} design configuration and is assumed parallel to the ground.} 
\label{fig:Fig1}} 
\end{figure}

The concept of serving and best panel for the MPUE architecture that is used in the measurement reporting for Tx beam management and HO procedure, respectively, were introduced in one of our earlier works \cite{b4}. This paper builds upon that and introduces the allied concept of serving and best Rx beam. In line with 3GPP \cite{b115}, the signal measurement scheme that we consider is MPUE-A3, where it is assumed that the UE can measure the RSRP values from the serving cell $c_0$ and neighboring cells by activating all three panels simultaneously. 3GPP defines UE-side Rx-beamforming as a follow-up procedure to the Tx serving beam selection procedure at the BS-side discussed in \Cref{Subsection2.2}. After the selection of a serving beam $b_0$ based on SSB, the UE can sweep through its beams and thereby select a narrow refined beam \cite[Chapter 4.2]{b5, b9}. Herein, the serving cell now repeats the channel state information reference signal (CSI-RS) associated with the serving beam $b_0$ for some time while the UE is sweeping its Rx beams on its panels. In our implementation, it is assumed that the serving beam on CSI-RS has the same beamwidth as the serving beam based on SSB. Furthermore, it is assumed that the Rx beam sweep for $b_0$ can be completed within the designated SSB period. Once the Rx beam sweep is complete, the UE adjusts and selects the beam with the highest L1 RSRP. The serving panel~$d_0$ and serving Rx beam $r_0$ are defined as
\vspace{-0.25\baselineskip}
\begin{equation}
\label{Eq7} 
[d_0, r_0] = \arg \max_{d,r} P_{c_0,b_0,d,r}^\mathrm{L1}(m).
\end{equation}
The serving panel and Rx beam selection decisions are both fully UE-centric and made independent of the network. As seen in \cite{b4}, the serving panel $d_0$ serves two key purposes. Firstly, it is used for beam reporting for intra-cell Tx beam management, as discussed in \Cref{Subsection2.2}. Secondly, the raw beam panel RSRPs measured on $d_0$ are used for calculating the average downlink SINR $\gamma_{c,b}$ of a link between the UE and beam $b$ of cell $c$. As will be discussed later in \Cref{Section4}, the SINR $\gamma_{c,b}$ is used for HOF and RLF determination. An illustration of the serving panel and serving Rx beam in the MPUE context can be seen Fig. \,\ref{fig:Fig1}.

The best beam~$r_c$ is chosen as the beam with the strongest L1 beam panel RSRP~$P_{c,b,d,r}^{\mathrm{L1}}(m)$ on the best panel~$d_c$ for any beam~$b$ of cell~$c$ in the network and is defined as
\vspace{-0.4\baselineskip}
\begin{equation}
\label{Eq8} 
[d_c, r_c] = \arg \max_{b,d,r} P_{c,b,d,r}^{\mathrm{L1}}(m).
\end{equation}
Herein, it is assumed that the UE can determine the best Rx beam with respect to the strongest L1 beam panel measurement of cell $c$. The L1 beam panel RSRPs $P_{c,b,d,r}^{\mathrm{L1}}(m)$ of the best panel $d_c$ are denoted as L1 beam RSRPs $P_{c,b}^{\mathrm{L1}}(m)$ and are used for deriving the L3 cell quality measurement $P_{c}^{\mathrm{L3}}(m)$ and L3 beam measurements $P_{c,b}^{\mathrm{L3}}(m)$, as explained in \Cref{Subsection2.1}.  These L3 cell quality measurements are then used for HO decisions. It is pertinent to mention here that the standard does not mandate that Rx-beamformed measurements are used in HO decision-making \cite{b8}. Therefore, the HO decision could also be made dependent on the wide Rx beam on the best panel $d_c$ and would not involve beam sweeping. As such, (\ref{Eq8}) reduces to (8) in \cite{b4}.

\begin{figure}[!t]
\textit{\centering
\includegraphics[width = 0.96\columnwidth]{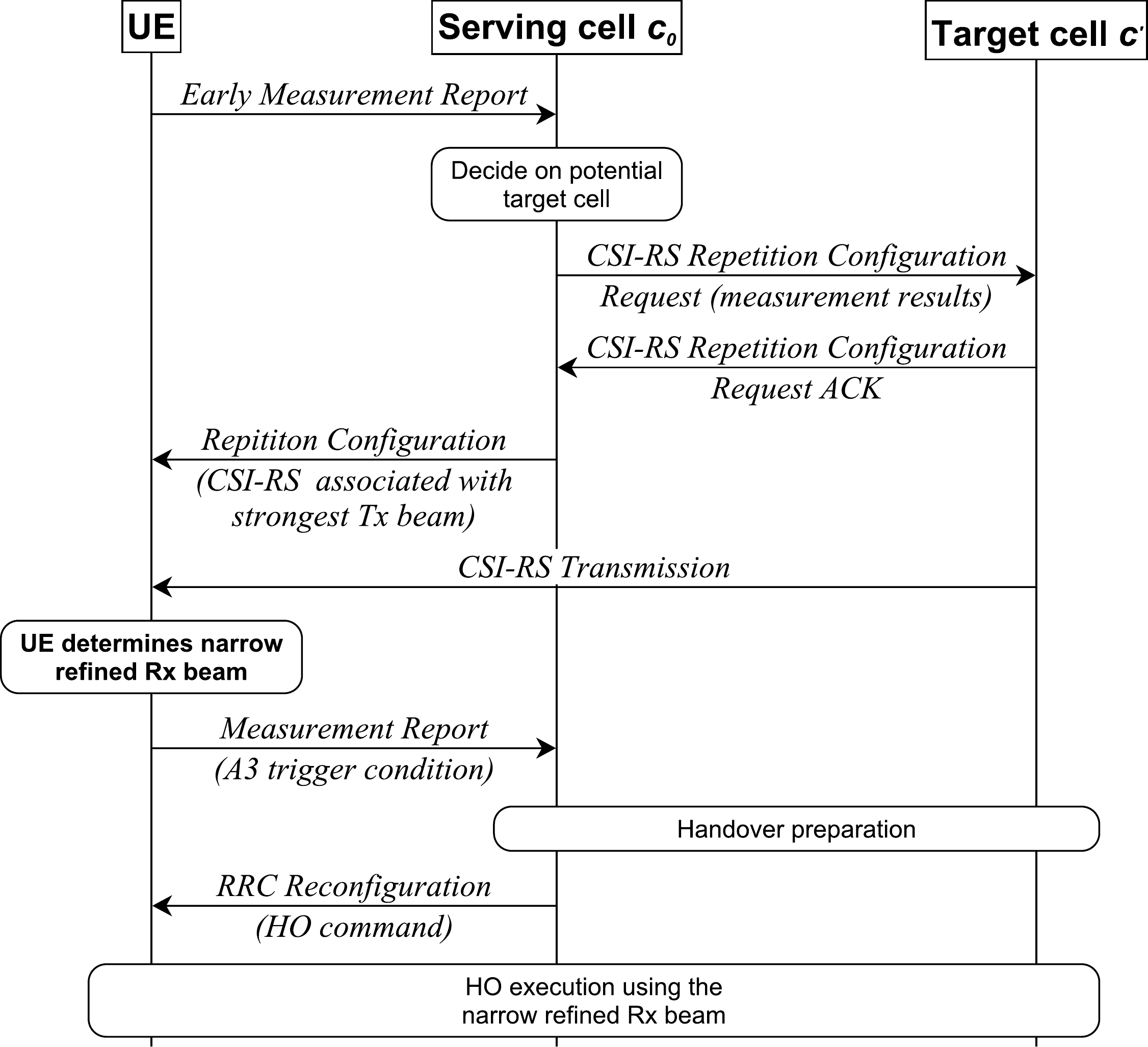}
\vspace{-2pt}
\caption{Signaling diagram showing the proposed enhancement \cite{b65} of acquiring the narrow refined Rx beam before a HO.} 
\vspace{-2pt}
\label{fig:Fig2-5}} 
\end{figure}

In the current standard \cite{b8}, UE-side beam refinement is only performed after a HO is complete. In \cite{b65}, an enhancement is proposed whereby the UE can acquire the narrow refined Rx beam of a target cell $c^{\prime}$ before a HO is initiated and use it during handover execution. This helps achieve a higher Rx-beamforming gain over the target link and less interference while performing the random access, thus enhancing the mobility performance in terms of HOFs. However, the implementation is far from trivial and the concept has not been validated and analyzed in a system-level simulation. In this paper, we implement this concept in our simulation framework and then validate the findings. The signaling diagram for the proposed scheme is illustrated in Fig.\,\ref{fig:Fig2-5}. Herein, it is seen that the UE sends an early measurement report to the serving cell~$c_0$ before a HO is initiated, which then decides on a potential target cell $c^{\prime}$. The serving cell then sends a \textit{CSI-RS Repetition Configuration Request} to the target cell, requesting a repetition of the CSI-RS associated with the Tx beam that has the strongest L1 beam RSRP in the measurement report. This is acknowledged by $c^{\prime}$. The serving cell then forwards the \textit{Repetition Configuration} message to the UE, after which the target cell transmits the CSI-RS associated with its strongest Tx beam. Thereafter, the UE sweeps its beams and determines the narrow refined Rx beam on a panel as per the same concept as seen in (\ref{Eq7}). The UE then sends a measurement report based on the A3 trigger condition in (\ref{Eq6}). Having received the measurement report, the serving cell prepares the handover. Thereafter, the serving cell $c_0$ sends an \textit{RRC Reconfiguration} message to the UE, which is followed by a HO execution using the narrow refined Rx beam that has been acquired.

\vspace{-2mm}
\subsection{UE-Side Beamforming Approaches} \label{Subsection3.2}
The employment of UE-side Rx-beamforming in the MPUE architecture means that different approaches can be defined whereby Rx-beamforming is used in the system model. These approaches are summarized in \Cref{Table1}. The reference approach considers a single antenna element per panel~\cite{b4} and therefore there is no Rx-beamforming capability. In Rx-beamforming Approach~1, the effect of Rx-beamforming is considered and impacts both the Tx beam management and serving link SINR.  Rx-beamforming Approach 2 extends the concept of Rx-beamforming Approach 1 and incorporates the proposed enhancement of refined Rx target beam acquisition into the system model. Lastly, Rx-beamforming Approach~3 extends Rx-beamforming Approach~2 and now also incorporates the Rx-beamformed measurements based on the refined Rx beam into the L3-measurements-based HO decision.

\begin{table}[!t]
\begin{center}
\caption{Rx-beamforming approaches}
 \vspace{-4pt}
\label{Table1}
\begin{tabular}{| c | c | c | c |}
\hline
\bfseries Approach & \bfseries Tx Beam   & \bfseries Refined Rx   & \bfseries  L3 \\
\bfseries & \bfseries  Management \&   & \bfseries  Target Beam   & \bfseries HO  \\
\bfseries & \bfseries Serving Link & \bfseries Acquisition  & \bfseries  Decision \\
\hline
Reference & (No) Rx- & (No) Rx- & (No) Rx- \\
Approach  &  beamforming &  beamforming & beamforming \\
\hline
Rx-beamforming & Rx- & (No) Rx- & (No) Rx- \\
Approach 1 &  beamforming &  beamforming & beamforming \\
\hline
Rx-beamforming  & Rx- & Rx- & (No) Rx- \\
Approach 2 &  beamforming & beamforming &  beamforming\\
\hline
Rx-beamforming  & Rx- & Rx- & Rx- \\
Approach 3 &  beamforming & beamforming &  beamforming\\
\hline
\end{tabular}
\end{center}
 \vspace{-8pt}
\end{table}

%
\section{Simulation Scenario and Parameters} \label{Section4}
In this section, the simulation setup for the 5G network model is discussed along with the simulation parameters that are listed in \Cref{Table2}. The simulations have been performed in our proprietary MATLAB-based system-level simulator. 

\begin{table}[!t]
\renewcommand{\arraystretch}{1.3}
\caption{Simulation parameters}
\centering
\begin{tabular}{|l | l|}
\hline
\bfseries Parameter & \bfseries Value\\
\hline\hline
Carrier frequency $f_\mathrm{FR2}$ & \SI{28}{GHz}\\
\hline
System bandwidth & \SI{100}{MHz}\\
\hline
Cell deployment topology & 7-site hexagon\\
\hline
Total number of cells $N_\mathrm{cells}$ & 21 \\
\hline
Downlink Tx power & \SI{40}{dBm}\\
\hline
Tx (BS) antenna height & \SI{10}{m}\\
\hline
Tx antenna element pattern & Table 7.3-1 in \cite{b12} \\
\hline
Tx panel size & 16 $\times$ 8, $\forall b \in~\{1,\ldots,8\}$\\
 & 8 $\times$ 4, $\forall b \in~\{9,\ldots,12\}$ \\
 \hline
Tx antenna element spacing & vertical: 0.7$\lambda$\\ 
& horizontal: 0.5$\lambda$\\
\hline
Tx beam elevation angle $\theta_b$ & 90$^{\circ}$, $\forall b \in~\{1,\ldots,8\}$ \\
 & 97$^{\circ}$, $\forall b \in~\{9,\ldots,12\}$\\
 \hline
Tx beam azimuth angle $\phi_b$ & $-$52.5$^{\circ}$$+$15$(b-1)^{\circ}, \forall b \in~\{1,\ldots,8\}$\\
& $-$45$^{\circ}$$+$30$(b-9)^{\circ}, \forall b \in~\{9,\ldots,12\}$\\
\hline
Tx-side beamforming model & Fitting model of \cite{b15}\\
\hline
Rx (UE) antenna height & \SI{1.5}{m}\\
\hline
Rx antenna element pattern & Table A.2.1-8 in \cite{b5}\\ 
\hline
Rx panel size  & Reference approach: 1 
\\ 
 & Rx-beamforming approaches: \\
 & 1 $\times$  4 $\forall r \in \{1,\ldots,7\}$\\
\hline
Rx antenna element spacing & horizontal: 0.5$\lambda$\\
\hline
Rx beam elevation angle $\theta_r$ & 90$^{\circ}$, $\forall r~\in \{1,\ldots,7\}$ \\
\hline
Rx beam azimuth angle $\phi_r$ & $-$45$^{\circ}$$+$15$(r-1)^{\circ}, \forall r \in \{1,\ldots,7\}$\\
\hline
Total number of UEs $N_\mathrm{UE}$ & 420\\
\hline
UE speed  & \SI{60}{km/h}\\
\hline
Fast-fading channel model & Abstract model of \cite{b15}\\
\hline
Number of simultaneously   & 4\\
scheduled beams per cell $K_b$ & \\ 
\hline
HO offset $o^\mathrm{A3}_{c_0, c^{\prime}}$  & \SI{2}{dB}\\
\hline
HO time-to-trigger $T_\mathrm{TTT,A3}$ & \SI{80}{ms}\\
\hline
Time step $\Delta t$  & \SI{10}{ms}\\
\hline
SSB periodicity & \SI{20}{ms}\\
\hline
Normalized L1 measurement  & 2\\
period $\omega$ &  \\
\hline
L1 filter length $N_\mathrm{L1}$ & 2\\
\hline
Simulated time $t_\mathrm{sim}$ & \SI{30}{s} \\
\hline
SINR threshold $\gamma_\mathrm{out}$  & \SI{-8}{dB} \\
\hline
\end{tabular}
\label{Table2}
\end{table}

\begin{figure}[!t]
\vspace{-\baselineskip}
\textit{\centering
\includegraphics[width = 0.96\columnwidth]{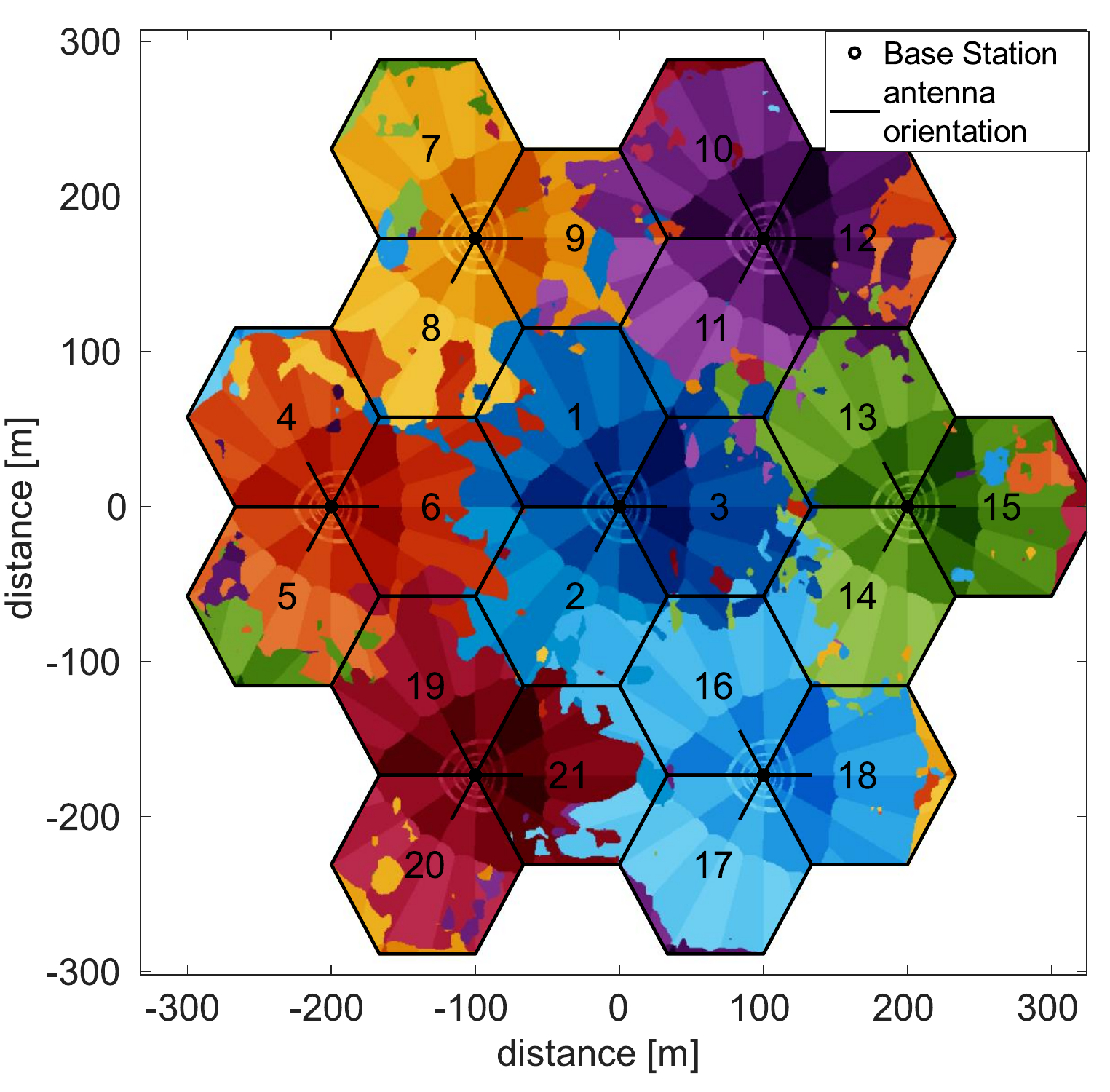}
\caption{Simulation scenario consisting of seven hexagonal sites, where each site is serving three cells with 120$^{\circ}$ coverage. } 
\label{fig:Fig3}} 
\end{figure}

We consider a 5G network model with an urban-micro (UMi) cellular deployment consisting of a standard hexagonal grid with seven BS sites, each divided into three sectors or cells. The inter-cell distance is 200 meters and the FR2 carrier frequency is \SI{28}{GHz}. 420 UEs are dropped randomly following a 2D uniform distribution over the network at the beginning of the simulation, moving at constant velocities along straight lines where the direction is selected randomly at the start of the simulation \cite[Table 7.8-5]{b12}. A wrap-around \cite[pp. 140]{b13} is considered, i.e., the hexagonal grid with seven BS sites is repeated around the original hexagonal grid shown in Fig.\,\ref{fig:Fig3} in the form of six replicas. This implies that the cells on network borders are subject to interference from the network's other edge that is comparable to those not on the network borders. All the UEs travel at 60 km/h, which is the usual speed in the non-residential urban areas of cities \cite{b14}.

In accordance with 3GPP's study outlined in \textit{Release 15} \cite{b12}, the channel model considered in this article takes into account shadow fading due to large obstacles  and assumes a soft line-of-sight (LoS) for all radio links between the UEs and the cells. Soft LoS is defined as a weighted average of the LoS and non-LoS channel components \cite[pp. 59-60]{b12} and is used for both shadow fading and distance-dependent path loss calculation. We take fast fading into account through the low complexity channel model for multi-beam systems proposed in \cite{b15}, which integrates the spatial and temporal characteristics of 3GPP's geometry-based stochastic channel model \cite{b12} into Jake’s channel model. The Tx-side beamforming gain model is based on \cite{b15}, where a 12-beam grid configuration is considered. Beams $b \in \{1,\ldots,8\}$ have smaller beamwidth and higher beamforming gain and cover regions further apart from the BS. Tx beams  $b \in \{9,\ldots,12\}$ have larger beamwidth and relatively smaller beamforming gain and cover regions closer to the BS. This can also be seen in Fig.\,\ref{fig:Fig3}, where the eight outer beams are shown in light colors and the four inner beams are shown in dark colors. The number of simultaneously scheduled beams per cell is taken as $K_b=$ 4. 

The antenna element radiation pattern for each of the three MPUE panels is based on \cite{b5}. The four antenna element patterns per panel produce seven Rx beams $\forall r \in \{1,\ldots,7\}$, where the elevation angle is $\theta_r=$~ 90$^{\circ}$ and the azimuth angle is $\phi_r=$~$-$45$^{\circ}$$+$15$(r-1)^{\circ}$. The UE screen, held by the user, is assumed to be parallel to the ground \cite{b4}. 

As mentioned in \Cref{Subsection2.3}, the SINR-dependent HOF and RLF models are now discussed below.

\textit{HOF Model:} A HOF is a failure over the target link that models the failure of a UE to handover from its serving cell $c_0$ to its target cell $c^{\prime}$. The UE initiates a handover by using the CFRA resources to access the selected beam $b^{\prime}$ of target cell $c^{\prime}$. For successful random access, it is a prerequisite that the SINR $\gamma_{c^{\prime},b^{\prime}}(m)$ of the target cell remains above the threshold $\gamma_\mathrm{out}$ during the RACH attempt, which is made after every 10 ms. A HOF timer $T_\mathrm{HOF} $ = \SI{200}{ms} is started when the UE initiates the random access towards the target cell $c^{\prime}$ and sends the RACH preamble. The RACH procedure is repeated until either a successful RACH attempt is achieved or $T_\mathrm{HOF}$ expires. A UE only succeeds in accessing the target cell if the SINR  $\gamma_{c^{\prime},b^{\prime}}(m)$ remains above the threshold $\gamma_\mathrm{out}$ and as such a successful HO is declared. A HOF is declared if the timer $T_\mathrm{HOF}$ expires and the UE fails to access the target cell, i.e., $\gamma_{c^{\prime},b^{\prime}}(m)<\gamma_\mathrm{out}$ for the entire duration that the HOF timer runs. The UE then performs connection re-establishment to a new cell (possibly the previous serving cell) and this procedure contributes to additional signaling overhead and signaling latency \cite{b7}. 

\textit{RLF Model:} An RLF is a failure over the serving link that models the failure of a UE while it is in the serving cell $c_0$. The UE further averages the average downlink SINR measurements of the serving cell $\gamma_{c_0,b_0}$ to yield the radio link monitoring (RLM) SINR metric, which it constantly keeps track of. An RLF timer $T_\mathrm{RLF}$ = \SI{1000}{ms} is started when the RLM SINR $\bar{\gamma}_\mathrm{RLM}$ of the serving cell $c_0$ drops below $\gamma_\mathrm{out} = $\SI{-8}{dB}, and if the timer $T_\mathrm{RLF}$ expires, an RLF is declared. The UE then initiates connection re-establishment. While the timer $T_\mathrm{RLF}$ runs, the UE may recover before declaring an RLF if the SINR $\bar{\gamma}_\mathrm{RLM}$ exceeds a second SINR threshold defined as $\gamma_\mathrm{in}$ = \SI{-6}{dB}, where $\gamma_\mathrm{in} > \gamma_\mathrm{out}$ \cite{b7}. As discussed in \Cref{Subsection2.2}, if the BFR process fails the UE also declares an RLF and this is also taken into account in the RLF model.

%
\section{Performance Evaluation}  \label{Section5}
In this section, the mobility performance of the reference approach is compared with the three different Rx-beamforming approaches. The mobility KPIs used for evaluation are explained below.

\subsection{KPIs} \label{Subsection5.1}
\begin{itemize}
    \item 
    \textit{RLFs:} Sum of the total number~of RLFs in the network.
    \item 
    \textit{HOFs:} Sum of the total number of HOFs in the network.
    \item 
    \textit{Successful HOs:} Sum of the total number of successful HOs from the serving  to the target cells in the network.
    \item 
    \textit{Fast HOs:} Sum of the total number of ping-pongs and short-stays in the network. A ping-pong is characterized as a successful HO followed by a HO back to the original cell within a very short time $T_{FH}$ \cite{b155}, e.g,. 1 second. It is assumed that both HOs could have been potentially avoided. A short-stay is characterized as a HO from one cell to another and then to a third one within $T_{FH}$. Here it is assumed that a direct HO from the first cell to the third one would have served the purpose. Although fast HOs are part of successful HOs, they are accounted for as a detrimental mobility KPI which adds unnecessary signaling overhead to the network.  

   \end{itemize}
   RLFs, HOFs, successful HOs, and Fast HOs are normalized to $N_\mathrm{UE}$ in the network per minute and expressed as UE/min.
   \begin{itemize}
     \item
    \textit{Total Outage:} Outage is denoted as a time period when a UE is not able to receive data from the network. This could be due to a number of reasons. When the SINR of the serving cell $\gamma_{c_0, b_0}$ falls below $\gamma_\mathrm{out}$ it is assumed that the UE is not able to communicate with the network and, thus, in outage. This is characterized as \textit{outage due to SINR degradation}. This outage type always precedes an RLF but it could also be that the SINR recovers before the RLF timer $T_\mathrm{RLF}$ expires. Besides, if the HOF timer $T_{\mathrm{HOF}}$ expires due to a HOF or the RLF timer $T_{\mathrm{RLF}}$ expires due to an RLF, the UE initiates connection re-establishment and this is also accounted for as outage. A successful HO, although a necessary mobility procedure, contributes also to outage since the UE cannot receive any data during the time duration the UE is performing random access to the target cell~$c^{\prime}$. This outage is modeled as relatively smaller (55ms) than the outage due to connection re-establishment (180 ms) ~\cite{b155}. The total outage in the network is denoted in terms of a percentage as 
\begin{equation}
\label{Eq10} 
\textrm{Total Outage} \ (\%) = \frac{\sum_{u}{\textrm{Outage duration of UE}} \ u} {N_\mathrm{UE} \ \cdot \ \textrm{Simulated  time}} \ \cdot \ 100. 
\end{equation}
\end{itemize}

%
\subsection{Simulation Results} \label{Subsection5.2}
\begin{figure}[!t]
  \begin{subfigure}{0.49\columnwidth}
  \includegraphics[width=\textwidth]{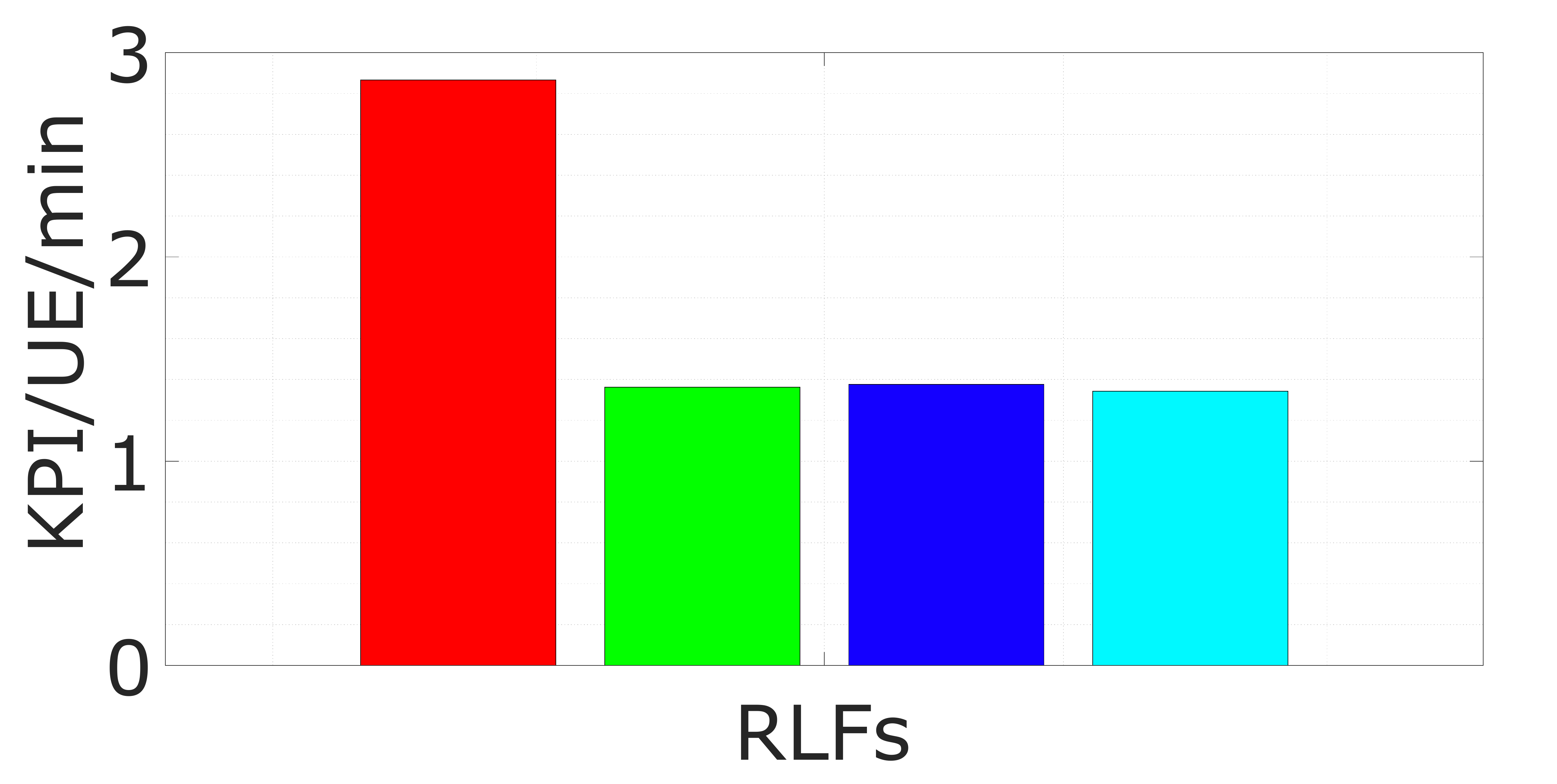}
  \vspace{-1.45\baselineskip}
  \caption{Number of RLFs.}
  \label{fig:Fig4a}
  \end{subfigure}
  \hfill
  \begin{subfigure}{0.49\columnwidth}
  \includegraphics[width=\textwidth] {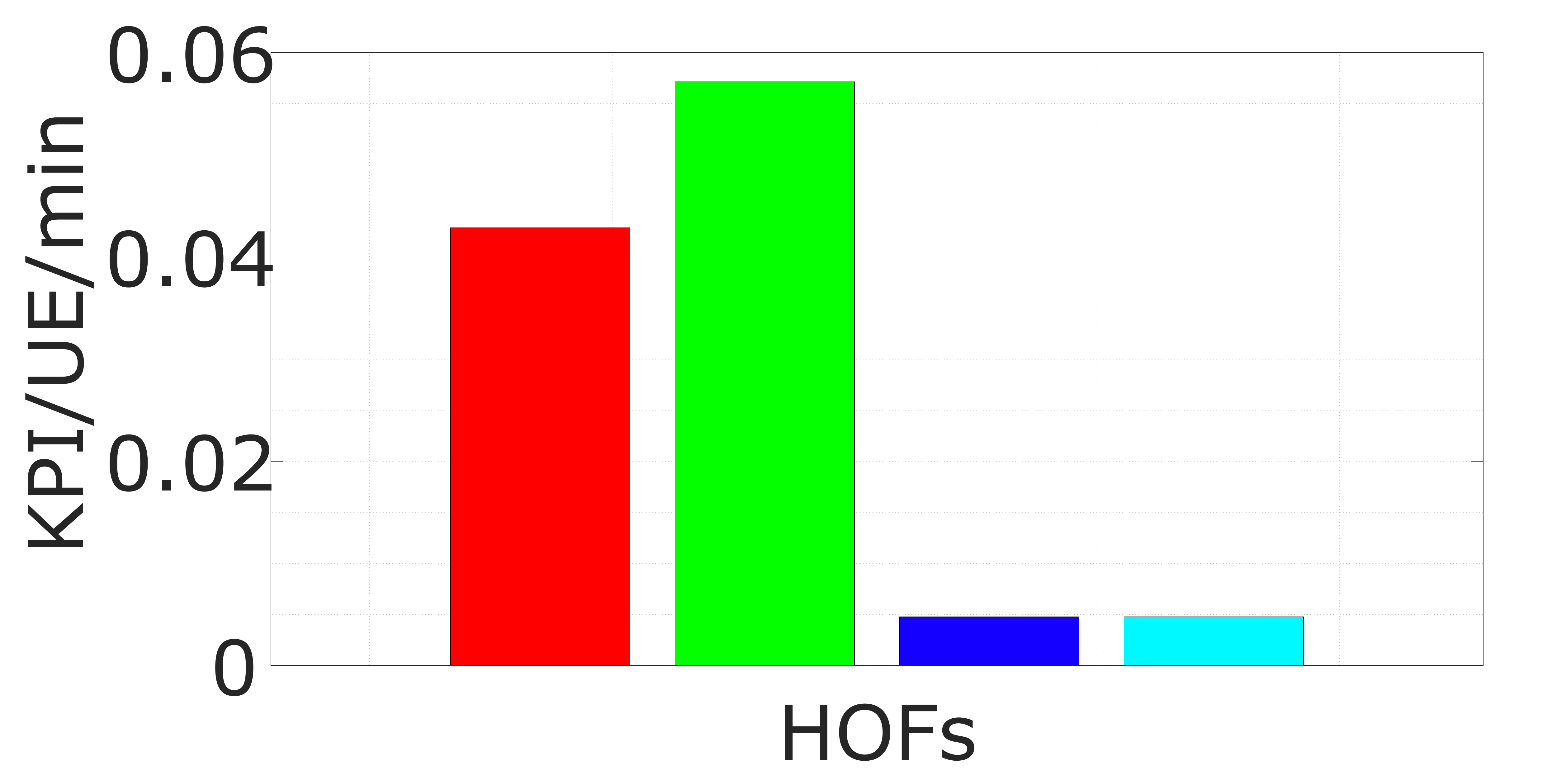}
  \vspace{-1.45\baselineskip}
  \caption{Number of HOFs.} 
  \label{fig:Fig4b}
  \end{subfigure} 
  \hfill
  \begin{subfigure}{0.45\columnwidth} 
  \includegraphics[width=\textwidth]{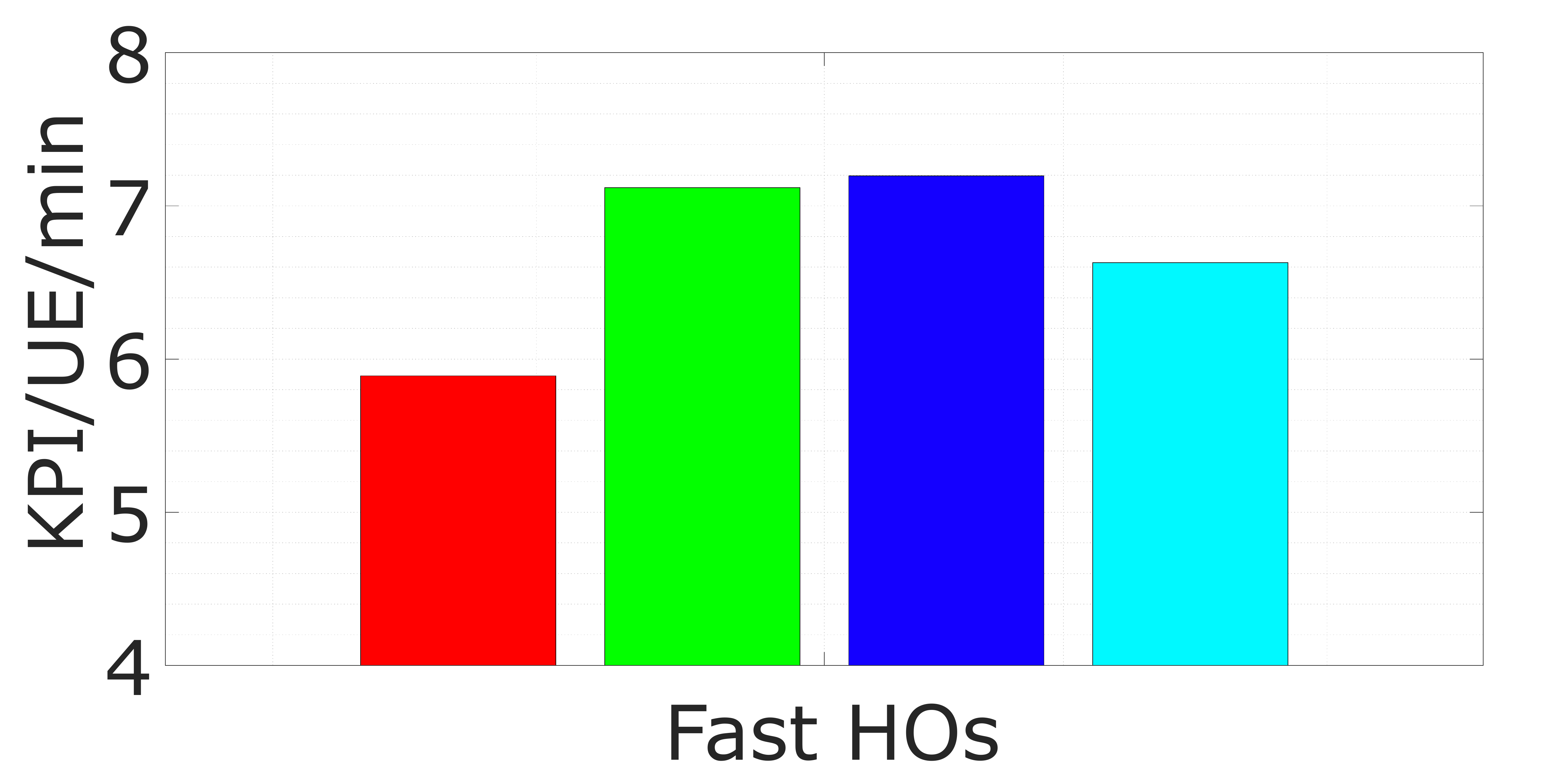} 
\vspace{-1.45\baselineskip}
  \caption{Number of fast HOs.} 
  \label{fig:Fig4c}
  \end{subfigure} 
  \hfill 
  \begin{subfigure}{0.45\columnwidth} 
  \includegraphics[width=\textwidth]{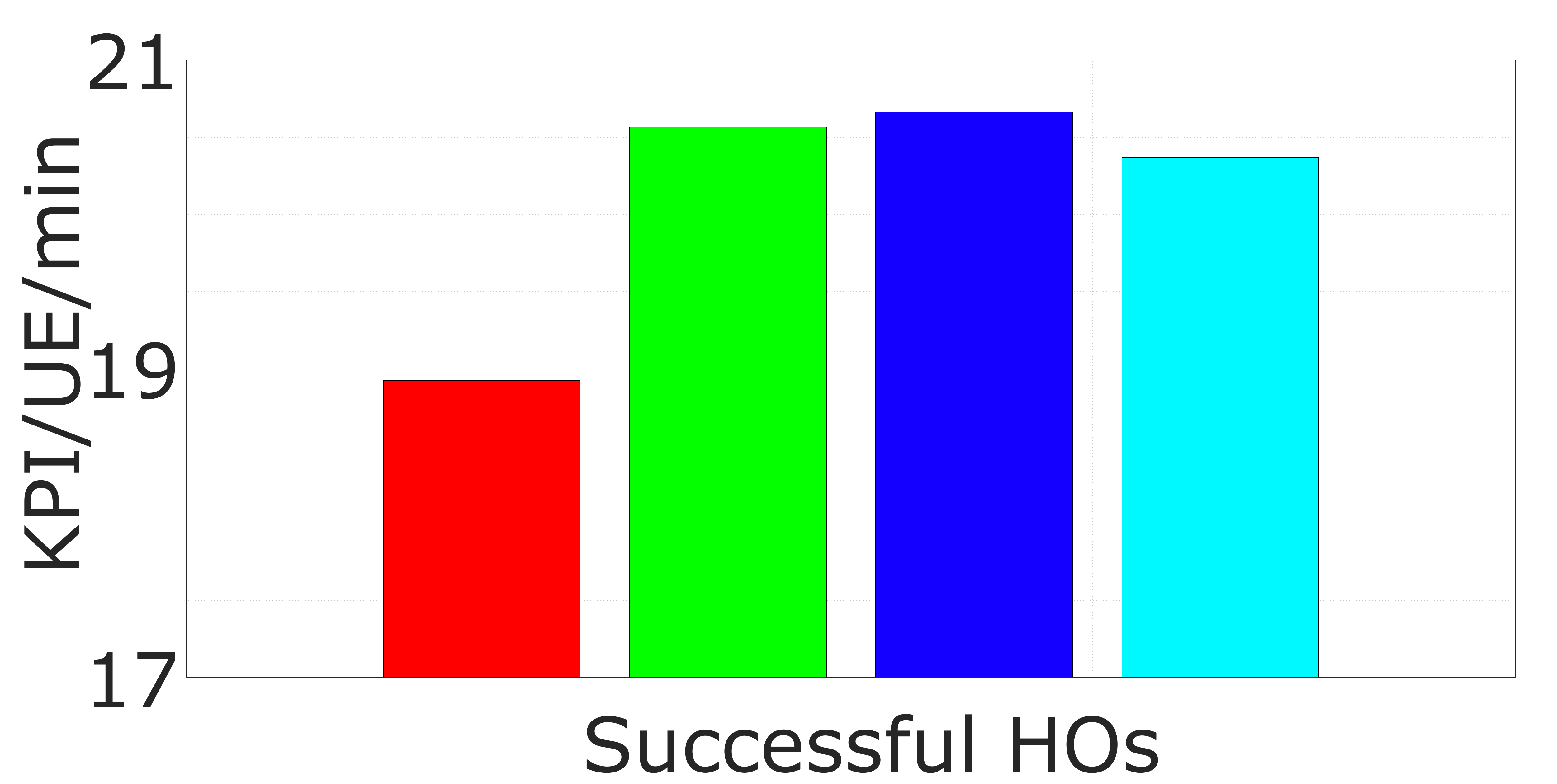} 
  \vspace{-1.45\baselineskip}
  \caption{Number of successful HOs.}
  \label{fig:Fig4d}
  \end{subfigure}
  \begin{subfigure}{0.5\textwidth}
  \centering
  \includegraphics[width=0.8\textwidth]{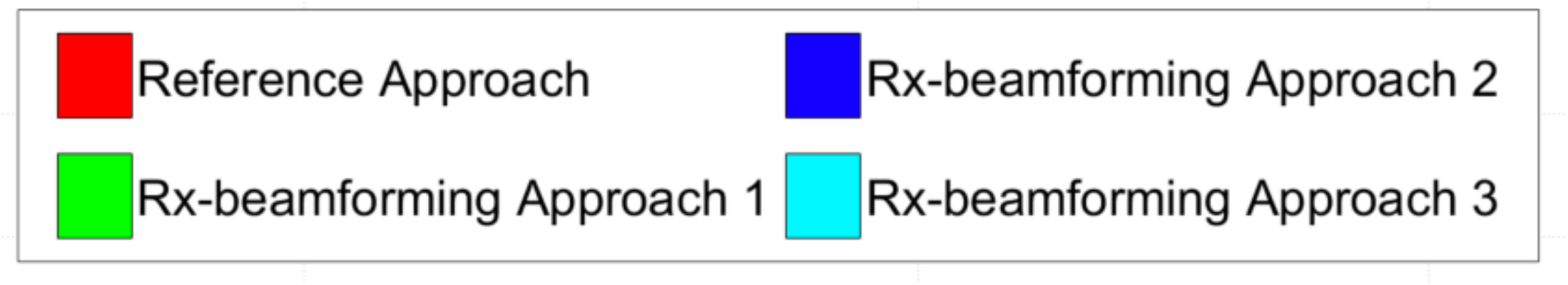} 
  \end{subfigure}
  \vspace{-16pt}
  \caption{A comparison of the mobility performance between the reference and the three different Rx-beamforming approaches in terms of total number (a) RLFs, (b) HOFs, (c) fast HOs, and (d) successful HOs.}
\label{fig:Fig4}
  \end{figure}

\begin{figure}[!b]
\textit{\centering
\includegraphics[width = 0.96\columnwidth]{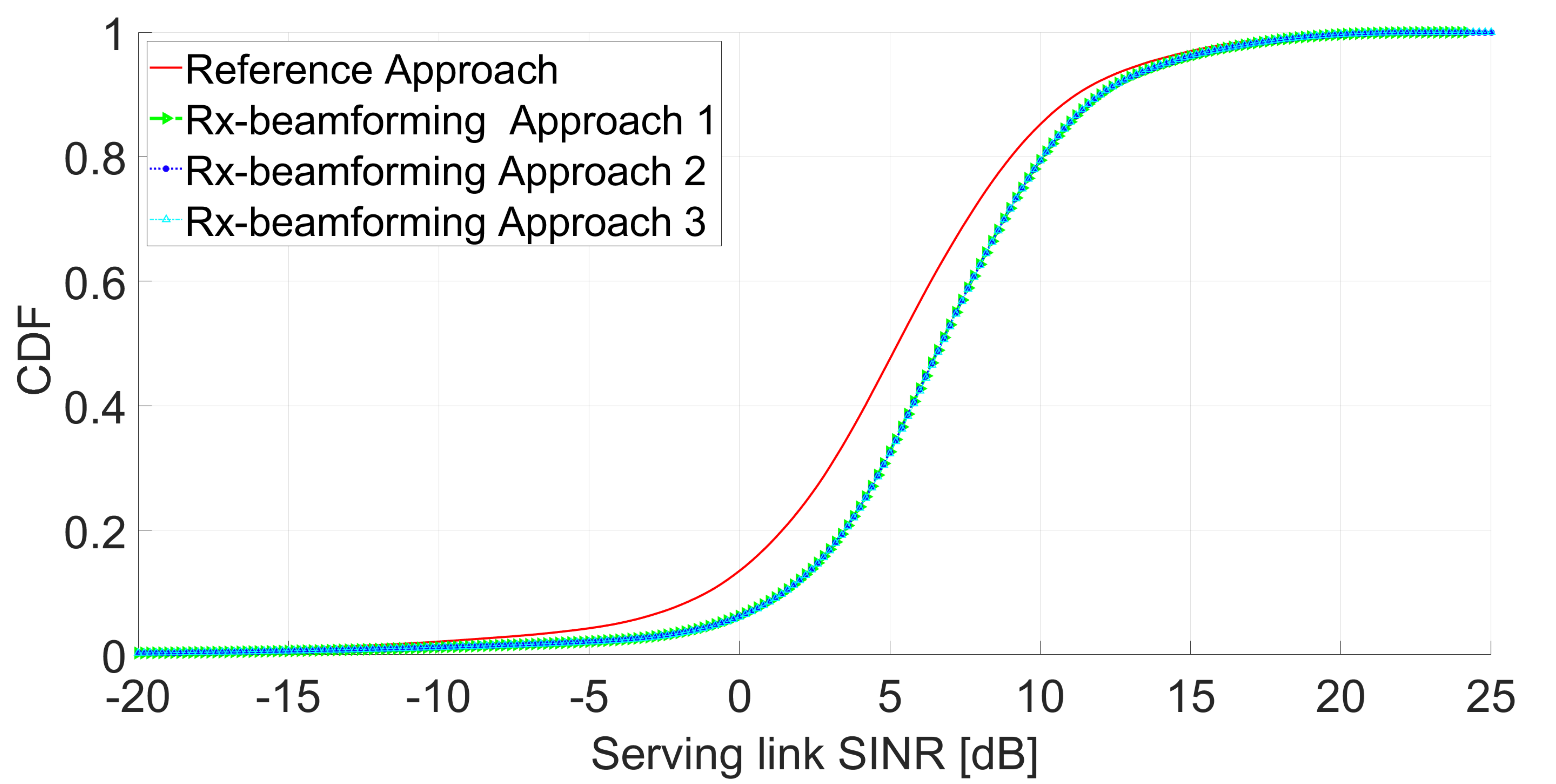}
\vspace{-4pt}
\caption{A comparison between the reference and the three different Rx-beamforming approaches in terms of the serving link SINR.} 
\label{fig:Fig5}} 
\end{figure}

Fig.\,\ref{fig:Fig4} shows a mobility performance comparison between the reference non-beamformed approach and the three different Rx-beamforming approaches in terms of RLFs, HOFs, fast HOs, and successful HOs. It is seen in Fig.\,\ref{fig:Fig4a} that for all the three Rx-beamforming approaches, there is an approximate 53\% relative reduction in RLFs when compared with the reference approach. This significant reduction stems from the fact that the serving link (and hence the serving link SINR $\gamma_{c_0, b_0}$) sees a significant improvement due to the Rx-beamformed measurements. This is illustrated in Fig.\,\ref{fig:Fig5}, where the CDF of the serving link SINR is shown. It is seen that at the 50\textsuperscript{th} percentile, the Rx-beamforming approaches (in blue, green, and cyan and overlapping) have an SINR of \SI{6.8}{dB} whereas the reference non-beamformed approach (shown in red) has an SINR of \SI{5.4}{dB}. Of particular interest is the low SINR regime in the vicinity of $\gamma_\mathrm{out}$ = \SI{-8}{dB}, since this is where mobility failures take place as explained by the HOF and RLF models in \Cref{Section4}. At the 2\textsuperscript{nd} percentile, the Rx-beamforming approaches have an SINR of \SI{-5.4}{dB} whereas the reference non-beamformed approach has an SINR of \SI{-10}{dB}. When the HOFs are analyzed in Fig.\,\ref{fig:Fig4b}, it is seen that for Rx-beamforming Approach 1 the HOFs increase by 32.6\% (from 0.043 HOFs/UE/min to 0.057 HOFs/UE/min) when compared with the reference approach. This performance degradation occurs because some RLFs may turn into HOFs because the UE makes more HO attempts as a result of the serving link SINR gain. With Rx-beamforming Approaches 2 and 3, the refined beam of the target link is acquired before the handover and the associated target link SINR gain significantly reduces HOFs by approximately 90\%. 

An analysis of the fast HOs in Fig.\,\ref{fig:Fig4c} reveals that as a consequence of the reduction in RLFs and HOFs, the UE attempts more HOs and a large number of them are fast HOs. For Approaches 1 and 2, the fast HOs increase relatively by 20\% when compared to the reference approach. With Approach 3, it is now seen that compared to the other two beamforming approaches, fast HOs can be curtailed by about 8\%. This is because the incorporation of the Rx-beamformed measurements into the L3 HO decision process makes it more reliable and therefore unnecessary HOs can be avoided. The same trend is also seen for successful HOs in Fig.\,\ref{fig:Fig4d}, where it is now seen that a reduction in RLFs and HOFs results in more successful HOs in general.

\begin{figure}[!b]
\textit{\centering
\includegraphics[width = 0.96\columnwidth]{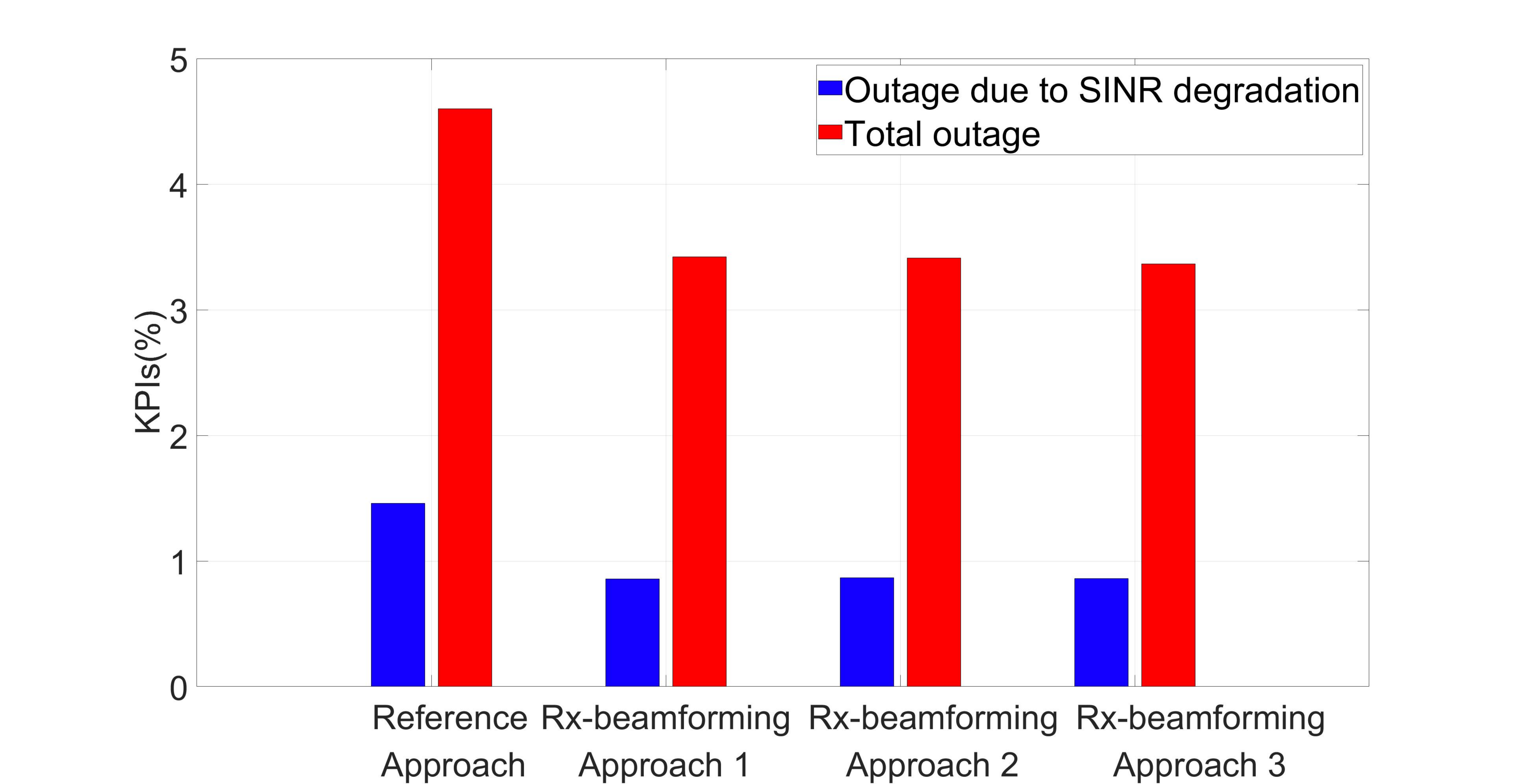}
\vspace{-2pt}
\caption{A comparison between the reference and the three different Rx-beamforming approaches for outage due to SINR degradation and total outage in the network.} 
\vspace{-18pt}
\label{fig:Fig6}} 
\end{figure}

Next, the outage performance is analysed in Fig.\,\ref{fig:Fig6}. It can be seen that for all the Rx-beamforming approaches, the outage due to SINR degradation (shown in blue) reduces by about 42\% in relative terms when compared with the reference approach. This is because the serving link SINR  $\gamma_{c_0, b_0}$ improves due to the beamforming gain as seen in Fig.\,\ref{fig:Fig5}. Due to this $\gamma_{c_0, b_0}$ falls below the SINR threshold $\gamma_\mathrm{out}$ less often. From our stimulative investigations, it is known that this is the second most common type of outage after outage due to successful handovers and therefore this is a significant improvement. It is also observed that the total outage (shown in red) reduces by 26\% in relative terms (from 4.60\% to 3.42\%) when Rx-beamforming Approach 1 is compared with the reference approach. This outage reduction stems from the reduction in outage due to SINR degradation and the outage reduction due to re-establishment as a result of less RLFs, as seen in Fig.\,\ref{fig:Fig4a}. It can also be observed that even though the outage due to successful HOs is increasing due to more successful HOs as seen in Fig.\,\ref{fig:Fig4d}, it is offset by the outage reduction due to re-establishment. This is because as mentioned in \Cref{Subsection5.1}, this outage is modeled as relatively smaller. With Rx-beamforming Approach 2, the total outage is comparable because the outage reduction due to a decrease in HOFs as seen in Fig.\,\ref{fig:Fig4b} is offset by the increased number of fast HOs (and therefore successful HOs) seen in Fig.\,\ref{fig:Fig4d}. There is a tradeoff between the HOFs and fast HOs in terms of their respective outage contribution. In mobility studies, it is known that minimizing the number of HOFs has a higher priority over fast HOs \cite{b155}. Lastly, it can be visualized that Approach 4 has the least total outage value of 3.36\% due to the reduction in fast HOs as seen in Fig.\,\ref{fig:Fig4d}, which results in less outage due to successful handovers.

\section{Conclusion} \label{Section6}
In this paper, the performance of Rx-beamforming for an MPUE architecture is analyzed in a multi-beam FR2 network. Both Rx-beamforming and MPUE are an integral part of 5G-Advanced and this paper is a novel attempt to understand how Rx-beamforming can help improve the serving link, the target link, and the L3 HO process. For this purpose, three different approaches are proposed. These approaches individually target the most common types of mobility problems, i.e,. RLFs, HOFs, and fast HOs. For a mobility setting based in non-residential areas of cities where the UE speed is assumed as 60 km/h, it is seen that compared to a non-beamformed reference setting, RLFs reduce by 53\% and HOFs reduce by 90\%. It is also seen that fast HOs increase due to a reduction in mobility failures and the use of Rx-beamformed measurements in the HO process can curtail them by 8\%. As a result, with Approach 3,  where all three mobility problems are targeted, it is seen that the outage reduces by 26\% when compared with the reference approach. Based on these findings, future studies may be carried out to investigate the mobility performance of Rx-beamforming with UE hand blockage effect \cite{b61, b16}.

%



\begin{thebibliography}{00}
\vspace{-2mm}
\bibitem{b1} S. Rangan, T. S. Rappaport, and E. Erkip, “Millimeter-Wave Cellular Wireless Networks: Potentials and Challenges,” \textit{Proceedings of the IEEE}, vol. 102, no. 3, pp. 366–385, 2014.
\bibitem{b2}  A. Ali et al., “System Model for Average Downlink SINR in 5G MultiBeam Networks,” \textit{in proc. of IEEE PIMRC}, 2019, pp. 1-6.
\bibitem{b3} F. Abinader, C. Rom, K. Pedersen, S. Hailu and N. Kolehmainen, “System-Level Analysis of mmWave 5G Systems with Different Multi-Panel Antenna Device Models,” \textit{2021 IEEE 93rd Vehicular Technology Conference (VTC2021-Spring)}, 2021, pp. 1-6.
\bibitem{b4} S. B. Iqbal, A. Awada, U. Karabulut, I. Viering, P. Schulz and G. P. Fettweis, “Analysis and Performance Evaluation of Mobility for Multi-Panel User Equipment in 5G Networks,” \textit{2022 IEEE 95th Vehicular Technology Conference (VTC2022-Spring)}, 2022, pp. 1-7.
\bibitem{b5} 3GPP, “Study on New Radio Access Technology Physical Layer Aspects (Release 14)”, TS 38.802 V14.2.0, 3GPP, Sep. 2017.
\bibitem{b6} 3GPP, “Evolution towards 5G-Advanced”, ETSI Webinar, May 2021, available at: https://www.3gpp.org/news-events/2194-ran webinar 2021.
\bibitem{b61} F. Fernandes, C. Rom, J. Harrebek, S. Svendsen and C. N. Manchón, “Hand Blockage Impact on 5G mmWave Beam Management Performance,” in \textit{IEEE Access}, vol. 10, pp. 106033-106049, 2022.
\bibitem{b62} A. Ali, J. Mo, B. L. Ng, V. Va and J. C. Zhang, “Orientation-Assisted Beam Management for Beyond 5G Systems,” in \textit{IEEE Access}, vol. 9, pp. 51832-51846, 2021, 
\bibitem{b65} A. Awada et al., \textit{User Equipment Beam Refinement Before Completion of Handover}, U.S. Patent Application 17/232,053, Oct. 2022. [Online]. Available: https://patents.google.com/patent/US20220338073A1/en.
\bibitem{b7} 3GPP, “NR; Radio Resource Control (RRC) Protocol Specification,” TS 38.331 V16.0.0, 3GPP, Mar. 2020.
\bibitem{b8} 3GPP, “NR; NR and NG-RAN Overall Description; Stage 2,” TS 38.300 V 16.4.0, 3GPP, Dec. 2020.
\bibitem{b9} M. Enescu, \textit{5G New Radio: A Beam-based Air Interface. Hoboken}, NJ, USA: Wiley 2020.
\bibitem{b10} 3GPP, “Requirements for Support of Radio Resource Management,” TS 38.133 V16.3.0, 3GPP, Mar. 2020. 
\bibitem{b115} 3GPP, \textit{Feature lead summary\#3 of Enhancements on Multi-beam Operations}, document R1-1907860, 3GPP TSG RAN WG1 Meeting \#97, LG Electronics, Reno, USA, May 2019.
\bibitem{b12} 3GPP, “Study on Channel Model for Frequencies from 0.5 to 100 GHz,” TR 38.901 V16.1.0, 3GPP, Dec. 2019.
\bibitem{b13} IEEE, \textit{802.16m Evaluation Methodology Document (EMD)}, IEEE 802.16 Broadband Wireless Access Working Group, Mar. 2008.
\bibitem{b14} 3GPP, “Technical Specification Group Radio Access Network; Study on Scenarios and Requirements for Next Generation Access Technologies;,” TR 38.913 V17.0.0, 3GPP, Mar. 2022.
\bibitem{b15}  U. Karabulut, A. Awada, I. Viering, A. N. Barreto and G. P. Fettweis, “Low Complexity Channel Model for Mobility Investigations in 5G Networks,” \textit{ in proc. of IEEE WCNC}, 2020, pp. 1-8.
\bibitem{b155} I. Viering, B. Wegmann, A. Lobinger, A. Awada and H. Martikainen, “Mobility robustness optimization beyond Doppler effect and WSS assumption,” \textit{2011 8th International Symposium on Wireless Communication Systems}, 2011, pp. 186-191.
\bibitem{b16} S. B. Iqbal, S. Nadaf, A. Awada, U. Karabulut, P. Schulz and G. P. Fettweis, "On the Analysis and Optimization of Fast Conditional Handover With Hand Blockage for Mobility," in \textit{IEEE Access}, vol. 11, pp. 30040-30056, 2023.

\end{thebibliography}
\end{document}